\documentclass[aps,preprint]{revtex4}%
\usepackage{amsfonts}
\usepackage{amsmath}
\usepackage{amssymb}
\usepackage{graphicx}%
\providecommand{\U}[1]{\protect\rule{.1in}{.1in}}
\begin{document}
\preprint{ }
\title[ ]{Comment on "Cornell potential in generalized uncertainty principle formalism:
the case of Schr\"{o}dinger equation"}
\author{Djamil Bouaziz}
\email{djamilbouaziz@univ-jijel.dz}
\affiliation{D\'{e}partement de Physique, Universit\'{e} de Jijel, BP 98, Ouled Aissa,
18000 Jijel, Algeria}
\begin{abstract}
In the recent paper \cite{has}, the $\ell$-waves Schr\"{o}dinger equation for
the Cornell's potential is solved in quantum mechanics with a generalized
uncertainty principle by following Ref. \cite{bou}. It is showed here that the
approach of Ref. \cite{bou} can only be used for the $s$-waves, and then the
solution given in \cite{has} would be true only in the special case $\ell=0.$
Furthermore, it is highlighted that the abstract and the conclusion of Ref.
\cite{has} do not accurately reflect the results of the paper.

\end{abstract}

\volumeyear{2016}
\startpage{1}
\maketitle

In Ref. \cite{has}, the authors considered the Cornell potential in quantum
mechanics with a generalized uncertainty principle (GUP). The Schr\"{o}dinger
equation for the $\ell-$waves has been written in momentum space (Eq. (9)),
and then, a quasi-exact analytical solution has been obtained for this
equation. To establish Eq. (9), the authors followed the approach of Ref.
\cite{bou}, used to study the hydrogen atom. However, this approach is only
valid for the s-waves and cannot be used when $\ell\neq0$. Indeed, it has been
shown in Ref. \cite{bou} that the definition of the operator $\widehat{R}$,
square root of the operator$\ \widehat{R}^{2}$, is only possible for the
$s$-waves in the case $\beta^{\prime}$ $=2\beta$, up to the first order of
$\beta$. The expressions of these operators are \cite{bou,kr}
\begin{align}
\widehat{R}^{2} &  \mathbf{=}\left(  i\hbar\right)  ^{2}\left\{  (1+6\beta
p^{2})\frac{d^{2}}{dp^{2}}+\frac{2}{p}\allowbreak(1+7\beta p^{2})\frac{d}%
{dp}\right\}  +O\left(  \beta^{2}\right)  ,\label{R1}\\
\widehat{R} &  \mathbf{=}i\hbar\left[  \left(  1+3\beta p^{2}\right)  \frac
{d}{dp}+\frac{1}{p}\allowbreak\allowbreak\left(  1+\beta p^{2}\right)
\right]  +O\left(  \beta^{2}\right)  .\label{R}%
\end{align}

It is to mention that the expression of $\widehat{R}^{2}$, as given in Ref.
\cite{has}, is not correct and cannot be used together with Eq. (\ref{R}) in
the case $\ell\neq0$, and then Eq. (9), which is the fundamental equation of
Ref. \cite{has}, is only true when $\ell=0$. Consequently, all the
transformations and parameters introduced to give the solution of Eq. (9) must
be independent of $\ell.$ Furthermore, the expression of $\widehat{R}$ was
presented without quoting Ref. \cite{bou}, where this result has been given
for the first time or Ref. \cite{kr}, where one can also find the formula of
$\widehat{R}$.

Moreover, the abstract of Ref. \cite{has} does not accurately reflect the
results of the paper. In fact, the claim "...as well as the set of equations
determining the spectrum of the system are obtained and the special case of
the vanishing minimal length parameter is recovered" is not strictly true.
First, there is no evident way to extract the energy spectrum from the
parameters defining the solution, and second, the limit $\beta=0$ has not been
correctly examined: instead of recovering the solution of ordinary quantum
mechanics ($\beta=0$) from the solution of Eq. (9), the authors written Eq.
(9) in the special case $\beta=0$, and then solved this equation once again.
Thereby, it has not been shown that the solution of Eq. (9) reduces to that of
the special case $\beta=0,$ which means that the correctness of the solution
is not checked.

\end{document}